\documentclass[prb, twocolumn, showpacs, showkeys,superscriptaddress]{revtex4}%
\usepackage{amsfonts}
\usepackage{amsmath}
\usepackage{amssymb}
\usepackage{psfrag}
\usepackage[dvips]{graphicx}%
\newcommand{\w}{\omega}

\begin{document}
\preprint{ }
\title{Electron transport in the four-lead two-impurity Kondo model:\\Nonequilibrium perturbation theory with almost degenerate levels}
\author{V. Koerting\footnote{Present address: Department of Physics, University of Basel, Klingelbergstrasse 82, CH-4056 Basel, Switzerland.} }
\email[author to whom correspondence should be addressed:\ ]
{verena.koerting@unibas.ch}
\affiliation{Institut f\"{u}r Theorie der Kondensierten Materie and DFG-Center for
Functional Nanostructures, Universit\"{a}t Karlsruhe, D-76128 Karlsruhe, Germany}
\author{J. Paaske}
\affiliation{The Niels Bohr Institute \& Nano-Science Center, University of Copenhagen,
DK-2100, Copenhagen, Denmark}
\author{P. W\"{o}lfle}
\affiliation{Institut f\"{u}r Theorie der Kondensierten Materie and DFG-Center for
Functional Nanostructures, Universit\"{a}t Karlsruhe, D-76128 Karlsruhe, Germany}


\begin{abstract}
The eigenstates of an isolated nanostructure may get mixed by the coupling to
external leads. This effect is the stronger, the smaller the level splitting
on the dot and the larger the broadening induced by the coupling to the leads
is. We describe how to calculate the nondiagonal density matrix of the
nanostructure efficiently in the cotunneling regime. 
As an example we consider a system of two quantum
dots in the Kondo regime, the two spins coupled by an antiferromagnetic
exchange interaction and each dot tunnel-coupled to two leads. 
Calculating the nonequilibrium density matrix and the corresponding current, we
demonstrate the importance of the off-diagonal terms in the presence of an
applied magnetic field and a finite bias-voltage.

\end{abstract}
\date{\today}
\maketitle

\section{Introduction}

In equilibrium, the occupation numbers, or more generally, the density matrix
of the quantum states of an interacting electron system as a function of
temperature $T$, chemical potential $\mu$ or magnetic field $B$, are given
explicitly in terms of the statistical operator. Out of equilibrium, even in a
stationary situation involving e.g. a current flow through the system, the
statistical operator is not known in general. Statistical expectation values
of observables may however be calculated for noninteracting systems, as e.g.
in the Landauer formula for the conductance, or else in perturbation theory in
the interaction using the method of nonequilibrium Green's
functions~\cite{Rammer:86, Haug:96}. 

A particularly simple example of the effect of a finite current on the
occupation of a quantum state is the polarization of a spin-1/2 on a quantum
dot in a magnetic field coupled by exchange interaction to the conduction
electron spins in the leads. For sufficiently large chemical potential
difference between source and drain electrodes, $\mu_{s}-\mu_{d}>g\mu_{B}B$,
where $g\mu_{B}B$ is the Zeeman splitting of the local spin levels, the
occupation numbers are no longer determined by the thermal Boltzmann factors,
but by a rate equation. More generally, it is a quantum Boltzmann equation
which describes the steady state of transitions between the two local Zeeman
levels induced by the available excess energy of electrons moving from the
reservoir at higher chemical potential (source) to that with lower chemical
potential (drain). These processes are mediated by the exchange interaction
and the resulting
occupation numbers, or equivalently, the spin polarization, can be very
different from their thermal equilibrium values, even in the limit of
vanishing exchange coupling. For example, the spin susceptibility, which in
equilibrium obeys the Curie law $\chi\propto1/T$, is found to decrease with
bias voltage $V$ as $\chi\propto1/V$ at $T\ll eV$ (cf.
Refs.~\onlinecite{Parcollet:02,Paaske:04}).

In the case of a single quantum dot characterized by a spin-1/2, the density
matrix of the local spin states is diagonal and the occupation numbers may be
obtained, at least in lowest order in the coupling, by solving a rate
equation. This is no longer the case in more complicated situations, when the
density matrix is not even approximately diagonal in the basis of eigenstates
of the isolated nanostructure or quantum impurity. 
The quantum Boltzmann equation therefore takes
the form of a matrix integral-equation which can easily become numerically
challenging. 

We show in this paper how the quantum Boltzmann equation may be
solved approximately in a controlled way 
with the aid of a non-unitary transformation of the 
matrix Green's functions for the quantum impurity. 
This transformation serves to diagonalize the impurity 
(matrix-)spectral-function. By subsequently neglecting the 
broadening of these eigenstates of the cotunnel-coupled impurity, 
one again arrives at a simple rate-equation for the occupation numbers, 
valid to leading order in the cotunneling amplitude (e.g.~second order 
in the exchange-coupling, i.e.~fourth order in the tunneling amplitudes). 
This Bloch-Redfield~\cite{Bloch:53, Redfield:57, Blum, Slichter} 
type equation still involves 
off-diagonal terms, describing the transition amplitudes for the 
(voltage-)driven impurity system, but it remains far simpler than 
the full quantum Boltzmann equation and is readily solved numerically. 
The Bloch-Redfield equations for the reduced density-matrix of a quantum dot 
have been employed in the sequential tunneling regime 
(cf.~e.g.~Refs.~\onlinecite{Engel,Lehmann}). 
Here we demonstrate how to establish similar equations in the cotunneling, 
or Kondo regime, starting from the quantum Boltzmann equation.

As an example of a system where the off-diagonal entries of the density matrix
become important, we study a system of two quantum dots in the Coulomb
blockade regime, each accommodating a single spin-1/2, mutually coupled by a
spin exchange interaction, $K$. Each dot is contacted by a set
of source and drain electrodes 
%ADD sentence
as illustrated in Fig.~\ref{fig:device}.
The eigenstates of the isolated two-impurity
system are the singlet and triplet states of the two coupled spins. 
Nevertheless, cotunneling via the leads may mix the states, whereby the
density matrix of the two-impurity system will acquire off-diagonal terms. As
demonstrated below, this mixing can only occur in the presence of an applied
magnetic field and for asymmetric couplings to the leads. 
The simpler case of zero magnetic field has already been analyzed in
a previous publication~\cite{Koerting:07}.
For $K = 0$, the product states of the two spin-1/2 would be a natural
basis, but for any finite $K$, perturbation theory should be performed
with respect to the singlet/triplet basis.
We will demonstrate how the singlet/triplet basis can be used, even
for $K = 0$, as long as the off-diagonal parts of the density
matrix are properly taken into account.

\section{Four-lead Two-impurity Kondo model}

\begin{figure}[t]
\centering
\includegraphics[width =0.9\columnwidth]{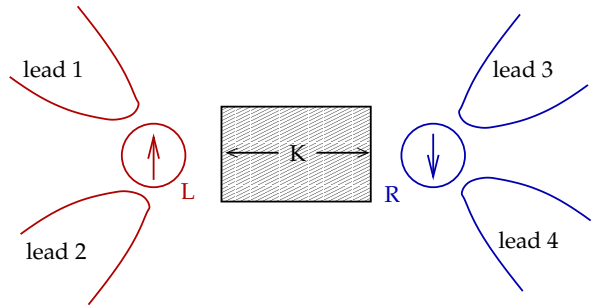} 
\caption{(color online) 
Double quantum dot setup: Two Kondo impurities, $L, R$, both
represented by a spin-1/2 are coupled mutually by a spin exchange interaction
$K$. The quantum dots are connected 
to two leads each, $1, 2$ and $3, 4$, respectively.}
\label{fig:device}%
\end{figure}

We model this four-lead two-impurity Kondo model as illustrated
in Fig.~\ref{fig:device} by the Hamiltonian
%OLD: The double dot system described above is modeled by the
%Hamiltonian
\begin{align}
H  &  =\sum_{\mathbf{k}n\sigma}(\epsilon_{k}-\mu_{n})
c_{n\mathbf{k}\sigma}^{\dagger}c_{n\mathbf{k}\sigma}\\
&  +K\mathbf{S}_{L}\cdot\mathbf{S}_{R}-g\mu_{B}\mathbf{B}\cdot
(\mathbf{S}_{L}+\mathbf{S}_{R})\nonumber\\
&  +{\textstyle\sum\limits_{n,m=1,2}}J_{L}^{nm}\,\mathbf{S}_{L}\cdot
\mathbf{s}_{nm}+{\textstyle\sum\limits_{n,m=3,4}}J_{R}^{nm}\,
\mathbf{S}_{R}\cdot\mathbf{s}_{nm},\nonumber
\end{align}
where $n,m=1,2,3,4$ labels the leads, which are characterized by the same
constant density of states near the Fermi level, $N(0)$, but with generally
different chemical potentials $\mu_{n}$. The spin-operator 
$\mathbf{s}_{nm}=\sum_{\mathbf{k},\mathbf{k}^{\prime},\sigma,\sigma^{\prime}}
c_{n\mathbf{k}\sigma}^{\dagger}(\vec{\tau}_{\sigma,\sigma\prime}/2)
c_{m\mathbf{k}^{\prime}\sigma^{\prime}}$ denotes respectively the
conduction electron spin ($n=m$) and the exchange tunneling operator ($n\neq
m$). Notice that we do not allow for charge transfer between the two dots. As
we have demonstrated earlier~\cite{Koerting:07}, the coupling to two pairs of
source, and drain electrodes gives rise to a marked transconductance signal,
reflecting the onset of cotunneling current through the left dot, say, when
tuning the voltage over the right dot to match the exchange coupling, i.e.
$\mu_{3}-\mu_{4}=K$.

The eigenstates of the isolated two-impurity spin system are the singlet and
triplet states
\begin{align*}
|t_{+}\rangle &  =|\uparrow\rangle_{L}|\uparrow\rangle_{R},\\
|t_{0}\rangle &  =\frac{1}{\sqrt{2}}\left(  |\uparrow\rangle_{L}%
|\downarrow\rangle_{R}+|\downarrow\rangle_{L}|\uparrow\rangle_{R}\right), \\
|t_{-}\rangle &  =|\downarrow\rangle_{L}|\downarrow\rangle_{R},\\
|s\rangle &  =\frac{1}{\sqrt{2}}\left(  |\uparrow\rangle_{L}|\downarrow
\rangle_{R}-|\downarrow\rangle_{L}|\uparrow\rangle_{R}\right).
\end{align*}
For $K=0$, however, it would be more reasonable to use the product states
\begin{align*}
|1\rangle &  =|\uparrow\rangle_{L}|\downarrow\rangle_{R}, \qquad
|2\rangle = |\downarrow\rangle_{L}|\uparrow\rangle_{R}, \\
|3\rangle &  =|\uparrow\rangle_{L}|\uparrow\rangle_{R}, \qquad
|4\rangle = |\downarrow\rangle_{L}|\downarrow\rangle_{R},
\end{align*}
of the left and the right spin as a basis. Expressing the latter basis in
terms of the former, the eigenstates with total spin quantum number $S_{z}=0$
are seen to mix:
\begin{align}
|1\rangle &  =|\uparrow\rangle_{L}|\downarrow\rangle_{R}
=\frac{1}{\sqrt{2}}\left(  |t_{0}\rangle+|s\rangle\right)  ,\nonumber\\
|2\rangle &  =|\downarrow\rangle_{L}|\uparrow\rangle_{R}
=\frac{1}{\sqrt{2}}\left(  |t_{0}\rangle-|s\rangle\right)  ,\label{eq:hyb}\\
|3\rangle &  =|t_{+}\rangle,\quad|4\rangle=|t_{-}\rangle.\nonumber
\end{align}

In order to have a convenient representation of the spin operators in the
singlet-triplet basis, we define a set of pseudo-boson (\textit{pb})
operators $\{b_{\gamma}^{\dagger}\}
=\{s^{\dagger},t_{+}^{\dagger},t_{0}^{\dagger},t_{-}^{\dagger}\}$, 
i.e.~$\gamma \in \{s,t_+,t_0,t_-\}$, to describe creation(annihilation) of a
singlet state $s^{\dag}$($s$) or a triplet state $t_{\gamma}^{\dag}%
$($t_{\gamma}$)~\cite{Sachdev:90}. 
The operators $b_{\gamma}^{\dagger}$ span an infinite
dimensional Fock space, which has to be projected onto the physical Hilbert
space, in which only single occupancy is allowed, i.e.~$Q=s^{\dagger}%
s+t_{0}^{\dagger}t_{0}+t_{+}^{\dagger}t_{+}+t_{-}^{\dagger}t_{-}=1$. This
constraint is enforced by adding a term $\lambda Q$ to the Hamiltonian and
taking the limit $\lambda\rightarrow\infty$~\cite{Abrikosov:65} when
calculating physical observables. The energy eigenvalues of the four states
are
\[
\omega_{s}=-\frac{3}{4}K,\quad \omega_{t_0}=\frac{1}{4}K,\quad 
\omega_{t_\pm}=\frac{1}{4}K\mp B,
\]
and therefore
\[
K\ \mathbf{S}_{L}\cdot\mathbf{S}_{R}=\omega_{s}s^{\dag}s+\sum_{\gamma}%
\omega_{t_0}t_{\gamma}^{\dagger}t_{\gamma}.
\]
In terms of the pseudo bosons, the spin-1/2 operators at the left/right dot
are given by
\begin{align}
S_{L/R}^{z} &  =\frac{1}{2}(\pm s^{\dagger}t_{0}\pm t_{0}^{\dagger}%
s+t_{+}^{\dagger}t_{+}-t_{-}^{\dagger}t_{-})
\label{eq:def_Sz}, \\
S_{L/R}^{+}=(S_{L/R}^{-})^{\dagger} &  =\frac{1}{2}(\pm s^{\dagger}t_{-}\mp
t_{+}^{\dagger}s+t_{+}^{\dagger}t_{0}+t_{0}^{\dagger}t_{-})
\label{eq:def_S+}.
\end{align}
Or in compact notation, with $\alpha=L,R$,
\[
\mathbf{S}_{\alpha}=\frac{1}{2}\sum_{\gamma,\gamma^{\prime}}\ b_{\gamma
}^{\dagger}\mathbf{T}_{\alpha;\gamma\gamma^{\prime}}b_{\gamma^{\prime}},
\]
where $\mathbf{T}_{\alpha;\gamma\gamma^{\prime}}$ is a vector of three 4x4
matrices, $T^{x},T^{y},T^{z}$, defined by
\[
T_{L/R}^{z}=%
\begin{pmatrix}
0 & 0 & \pm1 & 0\\
0 & 1 & 0 & 0\\
\pm1 & 0 & 0 & 0\\
0 & 0 & 0 & -1
\end{pmatrix},\quad 
T_{L/R}^{+}=%
\begin{pmatrix}
0 & 0 & 0 & \pm1\\
\mp1 & 0 & 1 & 0\\
0 & 0 & 0 & 1\\
0 & 0 & 0 & 0
\end{pmatrix},
\]
where $T_{\alpha}^{+}=(T_{\alpha}^{-})^{\dag}=\frac{1}{\sqrt{2}}(T_{\alpha
}^{x}+iT_{\alpha}^{y})$.

It is worth noting that the representation of $\mathbf{S}_{L/R}$, 
Eqs.~\eqref{eq:def_Sz} and \eqref{eq:def_S+}, does not
include an $s^{\dag}s$ term, but only transition operators from a singlet to a
triplet state. This implies, for example, that exchange-tunneling current 
cannot pass through either of the two quantum dots 
if the two-impurity system is in
the singlet state. Interestingly, the excitation gap, $K$, to the
current-carrying triplet states can be overcome by a finite bias across either
of the two dots. This gives rise to a pronounced transconductance signal,
which we have investigated earlier in Ref.~\onlinecite{Koerting:07}. This work
was restricted to zero magnetic field and hence avoided the problem of
off-diagonal terms in the nonequilibrium correlation functions.

\section{Density matrix and occupation numbers}

Assuming the exchange-tunneling to be weak, i.e.~$N(0)J_{\alpha}^{nm}\ll 1$, we
shall determine the singlet-triplet occupation numbers by means of
nonequilibrium perturbation theory. We employ contour-ordered
Green's functions arranged in the matrix-form
\[
\mathbf{G}=%
\begin{pmatrix}
G^{<}+G^{r} & G^{>}\\
G^{<} & G^{>}-G^{r}%
\end{pmatrix}
,
\]
satisfying the Dyson equation
\[
\mathbf{G}_{0}^{-1}\mathbf{G}=\mathbf{1}+\mathbf{\Sigma G},
\]
where $\mathbf{G}$ and $\mathbf{G}_{0}$ are the dressed and bare
contour-ordered matrix Green's functions and 
%before $\Sigma$
$\mathbf{\Sigma}$ is the self energy.

Whereas the information on the energy spectrum, including any shifts by
external fields, is encoded in the retarded Green's function $G^{r}$, the
information on the thermodynamic state of the system, i.e.~the occupation of
the energy levels, is contained in $G^{<}.$ Out of equilibrium, these two
functions are not simply connected through a fluctuation-dissipation theorem
and instead one must solve one more component of the matrix Dyson equation,
\[
G_{0}^{-1}\,G^{<}=\Sigma^{r}G^{<}+\Sigma^{<}G^{a}.
\]
An equivalent equation is obtained by applying 
$\mathbf{G}_0^{-1}$ to the second time argument in $\mathbf{G}$ 
(cf. Ref.~\onlinecite{Haug:96}).
In contrast to the Dyson equation for $G^{r},$ this quantum Boltzmann equation
is a self-consistent equation for $G^{<}$, since $\Sigma^{<}$ also depends on
$G^{<}$ and since no unrenormalized $G^{<}_0$ is known for an isolated
quantum impurity. In the limit of vanishing coupling to the leads, $G^{<}$ is
found to be independent of that coupling. Therefore the occupation number
\[
n=i\int\frac{d\omega}{2\pi}\,G^{<}(\omega)
\]
can take a finite limiting value dependent on the system parameters
temperature, bias voltage and magnetic field in $0$th order in the coupling to
the leads (cf. also Refs.~\onlinecite{Parcollet:02,Paaske:04}).
%CHANGE: before \cite now \onlinecite

In the example of a two-impurity system, the \textit{pb} Green's functions
%CHANGE: in the example instead of in our example
form a matrix in the singlet-triplet (ST) basis 
$\gamma=\{s,t_{+},t_{0},t_{-}\}$, i.e.
%CHANGE: i.e. instead of e.g.
\[
\mathcal{G}^{r}=%
\begin{pmatrix}
G_{ss}^{r} & 0 & G_{st_{0}}^{r} & 0\\
0 & G_{t_{+}t_{+}}^{r} & 0 & 0\\
G_{t_{0}s}^{r} & 0 & G_{t_{0}t_{0}}^{r} & 0\\
0 & 0 & 0 & G_{t_{-}t_{-}}^{r}%
\end{pmatrix}
,
\]
where only the elements allowed by symmetry are shown.

Notice that in the absence of any coupling to the leads, but assuming coupling
to a heat bath, the quantum impurity is in thermodynamic equilibrium and the
Green's functions are given immediately as
\begin{align}
G_{\gamma\gamma}^{<, (0)}(\omega)  &  =-in_B(\omega)A_{\gamma\gamma}^{(0)}(\omega
)\label{eq:G_DQD_0_<},\\
G_{\gamma\gamma}^{>, (0)}(\omega)  &  =-i[n_B(\omega)+1]A_{\gamma\gamma}^{(0)}(\omega
)\label{eq:G_DQD_0_>},\\
G_{\gamma\gamma}^{r/a, (0)}(\omega)  &  =\frac{1}{\omega-\omega_{\gamma}-\lambda\pm
i\delta}\label{eq:G_DQD_0_ra},\\
A_{\gamma\gamma}^{(0)}(\omega)  &  =2\pi\delta(\omega-\omega_{\gamma}-\lambda) 
\label{eq:A_DQD_0},
\end{align}
%CONSISTENCY: use n_B instead of n
where $\omega_{\gamma}$ is the energy of the state $\gamma\in\{s,t_{+}%
,t_{0},t_{-}\}$ and $n_B(\omega)$ is the Bose distribution function. 
%OLD:
In the limit $\lambda\rightarrow\infty$ the Bose function at the position of the
spectral peak, $\omega=\omega_{\gamma}+\lambda,$  turns into a Boltzmann
factor $n_{\gamma}\varpropto e^{-\beta(\omega_{\gamma}+\lambda)}\ll 1$. 
The difference between the Bose and the Fermi statistics 
for the pseudo particles 
is then seen to vanish. Any term containing a product of two occupation numbers
$n_{\gamma}^{\lambda}n_{\gamma^{\prime}}^{\lambda}\propto e^{-2\beta\lambda}$
will be projected out at the end of the calculation.

\subsection{Pseudo-boson self energy}

The first order (in $N(0)J_{\alpha}^{nm}$) pseudo-boson self energy is
proportional to the spin polarization of the conduction electrons. It provides
only a correction to the g-factor of the local spin and will be neglected in
the following.

\begin{figure}[b]
\centering
\psfrag{b}{$\beta, \beta'$}
\psfrag{g}{$\gamma$}
\psfrag{g'}{$\gamma'$}
\includegraphics{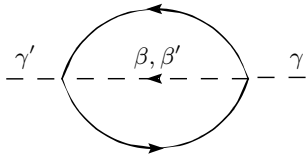}
\caption{
%IMPROVE: add legend to diagram
Diagram for the second order pseudo boson self energy. 
Solid lines: conduction electron Green's functions, dashed lines: pseudo
particles of the double quantum dot system.}
\label{fig:selfenergy}
\end{figure}
The second order self energy, on the other hand, corresponds to
the diagram in Fig.~\ref{fig:selfenergy} and reads
\begin{align}
\Sigma_{\gamma^{\prime}\gamma}(\tau_{1},\tau_{2})=-\frac{1}{16}  &
 \sum\limits_{\alpha=L,R}Y_{\alpha}(\tau_{1},\tau_{2}
)\label{eq:general_selfenergy_1}\\
&\times \sum_{\beta,\beta\prime}\mathbf{T}_{\alpha;\gamma^{\prime}\beta
}\ G_{\beta,\beta^{\prime}}(\tau_{1},\tau_{2}) \mathbf{T}
_{\alpha;\beta^{\prime}\gamma}.\nonumber
\end{align}
%CONSISTENCY: remove \vec in \vec{\mathbf{T}}
Here the time variables $\tau_{1,}\tau_{2}$ lie on the Schwinger contour, and
$\alpha=L(R)$ for $m,n=1,2,(3,4)$.
We have introduced the abbreviation,
\begin{equation}
Y_{\alpha}(\tau_{1},\tau_{2})=2\!\!\!\!\!\!\sum_{m,n=1,2,(3,4)}
J_{\alpha}^{mn}J_{\alpha}^{nm}\,X_{n}^{m}(\tau_{1},\tau_{2}),
\label{eq:YR_tau}
\end{equation}
with the summation variables depending on $\alpha$ and with
\[
X_{m}^{n}(\tau_{1},\tau_{2})=\frac{1}{(-i)^{2}}\sum_{k,k^{\prime}
}G_{nk^{\prime}}(\tau_{2},\tau_{1})G_{mk}(\tau_{1},\tau_{2})
\]
being the conduction electron susceptibility.

This expression is quite general and would be valid for any quantum impurity
with internal states $\gamma$, where the matrices $\mathbf{T}_{\alpha}^{i}$
would have to be defined accordingly. The self energy will have off-diagonal
components $(\gamma\neq\gamma\prime)$ if the conservation laws for spin allow
so. Since a conduction electron tunneling through one of the dots can either
flip its spin or not, the accessible intermediate states may change the
quantum number $S_{z}$ by $0,\pm 1$. The second process has to flip the
conduction electron spin back. Therefore, the self energy is diagonal in the
quantum number $S_{z}$. For the double dot considered here this leaves only
one possible off-diagonal element, $st_{0}$ (and its hermitian conjugate),
given by
\begin{align}
\Sigma_{st_{0}}(\tau_{1},\tau_{2})=  - \frac{1}{16} & 
\left(  G_{t_{0}s}(\tau_{1},\tau_{2})
[Y_{L}(\tau_{1},\tau_{2})+Y_{R}(\tau_{1},\tau_{2})]\nonumber \right.\\
&+[G_{t_{-}t_{-}}(\tau_{1},\tau_{2}) -G_{t_{+}t_{+}}(\tau_{1},\tau_{2})]\nonumber\\
&\left. 
 \hspace*{3mm}\times[Y_{L}(\tau_{1},\tau_{2})-Y_{R}(\tau_{1},\tau_{2})] \right).
\label{eq:Sigma_st0}
\end{align}
All other elements of the self energy are given in
appendix~\ref{sec:app_selfenergy}. We emphasize that the off-diagonal self
energy is finite only if two symmetries are broken simultaneously: time
reversal symmetry by a magnetic field $(t_{+}\not =t_{-})$ and parity, 
i.e.~the left-right symmetry $(Y_{L}\not =Y_{R})$.

\begin{figure}[b]
\centering
%\psfrag{s}{\footnotesize $s$}
\psfrag{t0}{\footnotesize $t_0$}
\psfrag{t+}{\footnotesize $t_+$}
\psfrag{t-}{\footnotesize $t_-$}
\includegraphics[width=0.4\textwidth]{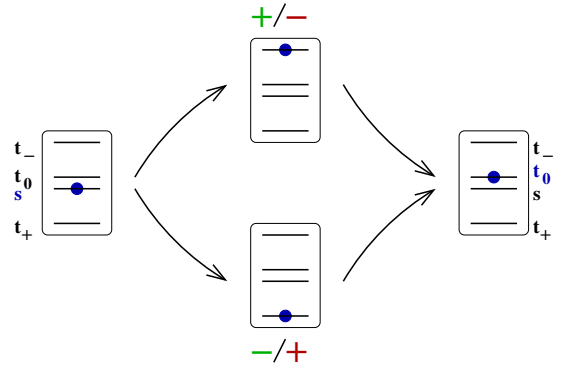}
\caption{(color online) 
Illustration of the symmetry in the off-diagonal self energy 
$\Sigma_{st_0}$. The two different paths from the
singlet-state $|s\rangle$ to the triplet-state $|t_{0}\rangle$ 
over $|t_-\rangle$ and $|t_+\rangle$ come with
opposite signs. Green(left) sign for interaction with left leads. Red(right)
sign for interaction with right leads.}
\label{fig:explain_st0}
\end{figure}This can be understood from the following simple argument.
Starting from the singlet state, a flipping of the left spin, say, causes a
transition to a triplet state with $S_{tot}^{z}\not =0$,
\begin{align}
S_{L}^{\mp}|s\rangle=\pm\frac{1}{\sqrt{2}}|t_{\mp}\rangle,
\end{align}
which, upon a subsequent flipping of the left spin, makes a transition to
either $|s\rangle$ or $|t_{0}\rangle$.
This is illustrated in Fig.~\ref{fig:explain_st0}. The two intermediate
triplet states come with opposite signs, i.e.~shifted in phase by $\pi$, and
in the case of zero magnetic field these two alternative paths from
$|s\rangle$ to $|t_{0}\rangle$ cancel and $\Sigma_{st_{0}}$ vanishes. This is
not the case for the diagonal component $\Sigma_{ss}$ in which the signs are
squared and cause no cancellation. The observed dependence on the left-right
symmetry, indicated in Fig.~\ref{fig:explain_st0}, arises in a similar way.

Performing the analytical continuation to the real-time axis, one finds after
Fourier-transformation in the relative time-variable,
\begin{align}
Y_{L}^{<}(\omega)=  &  (-2\pi)\left[  \left(  g_{11}^{2}+g_{22}^{2}\right)
2B(\omega)\right. \label{eq:YL_FT}\\
&  \left.  +2g_{12}g_{21}\left(  B(\omega+eV_{L})+B(\omega-eV_{L})\right)
\right] \nonumber,\\
Y_{R}^{<}(\omega)=  &  (-2\pi)\left[  \left(  g_{33}^{2}+g_{44}^{2}\right)
2\,B(\omega)\right. \label{eq:YR_FT}\\
&  \left.  +2g_{34}g_{43}\left(  B(\omega+eV_{R})+B(\omega-eV_{R})\right)
\right] \nonumber,
\end{align}
and $g_{nm}=N(0)J^{nm}, Y_{L\pm R}^{<}(\omega)=Y_{L}^{<}(\omega)\pm
Y_{R}^{<}(\omega)$, where we have introduced the function
\[
B(x)=x\cdot n_{B}(x)=\left[  \coth(\beta x/2)-1\right]  x/2.
\]
From these correlation functions, the lesser component and the imaginary part
of the retarded self energy are obtained as
\begin{align}
\Sigma_{st_{0}}^{<}(\omega)  &=- \frac{1}{16}
\int\frac{d\epsilon}{2\pi}\,
\left\{ G_{t_{0}s}^{<}(\epsilon)Y_{L+R}^{<}(\omega-\epsilon) \right.\nonumber\\
&  \qquad\qquad+ \left.
\left(  G_{t_{-}t_{-}}^{<}(\epsilon)-G_{t_{+}t_{+}}^{<}(\epsilon)\right)
Y_{L-R}^{<}(\omega-\epsilon) 
\right\}
\label{eq:Sigma_st0_FT},\\
\Gamma_{st_{0}}(\omega)  &  =2i\mathrm{Im}\Sigma_{st_{0}}^{r}\approx- 2
\frac{1}{16}\int\frac{d\epsilon}{2\pi}\{\,A_{t_{0}s}(\epsilon)Y_{L+R}^{<}
(\epsilon-\omega)\nonumber\\
&  \qquad\qquad+\left(  A_{t_{-}t_{-}}(\epsilon)-A_{t_{+}t_{+}}(\epsilon
)\right)  Y_{L-R}^{<}(\epsilon-\omega)\} \label{eq:Gamma_st0_FT}.
\end{align}
We observe that the off-diagonal spectral function obeys the sum rule
\[
\int\frac{d\omega}{2\pi}A_{st_{0}}(\omega) =
\left\langle [s,t_{0}^{\dag}]\right\rangle = 0 .
\]

From the hermiticity condition it follows that
\begin{equation}
\left[  G_{st_{0}}^{r}(\omega)\right]  ^{\ast}=G_{t_{0}s}^{a}(\omega)
\label{eq:rule_for_retarded_gst0},
\end{equation}
and thus the spectral functions,
\[
A_{st_{0}}(\omega)=A_{t_{0}s}(\omega),
\]
are identical. In addition one has
\begin{align}
&\left[  G_{st_{0}}^{<}(t)\right]^{\ast} = - G_{t_{0}s}^{<}(-t) \nonumber\\
\Rightarrow\qquad &\left\{
\begin{matrix}
\mathrm{Re}\left[  G_{st_{0}}^{<}(\omega)\right]   
=-\mathrm{Re}\left[  G_{t_{0}s}^{<}(\omega)\right]  \\ %\qquad\nonumber\\
\mathrm{Im}\left[  G_{st_{0}}^{<}(\omega)\right]   
=\mathrm{Im}\left[  G_{t_{0}s}^{<}(\omega)\right]
\end{matrix} \right.
  \label{eq:rule_for_lesser_gst0}
\end{align}
The diagonal elements of $G_{\gamma\gamma\prime}^{<}$ are purely imaginary
functions, and the off-diagonal elements share this property approximately
\begin{align}
\mathrm{Re}\left[  G_{st_{0}}^{<}(\omega)\right]   &  =-\mathrm{Re}\left[
G_{t_{0}s}^{<}(\omega)\right]  \approx0\nonumber\\
\Rightarrow G_{st_{0}}^{<}(\omega) &  \approx G_{t_{0}s}^{<}(\omega
)\label{eq:assum_Gst0_im}
\end{align}
Thus the lesser Green's function is assumed to be symmetric in analogy to the
spectral function which was proven to be symmetric. It follows
straightforwardly that $\Sigma_{st_{0}}=\Sigma_{t_{0}s}$.

\subsection{Retarded Green's function}

Assuming that the off-diagonal self energy is finite, we find by solving
Dyson's equation $\mathcal{G}^{r}=[(\mathcal{G}^{r, (0)})^{-1}-\Sigma^{r}]^{-1}$,
\[
\mathcal{G}^{r}=\frac{1}{\mathrm{det}}
\begin{pmatrix}
(G_{t_{0}t_{0}}^{r})^{-1} & 0 & \Sigma_{st_{0}}^{r} & 0\\
0 & \mathrm{det}\cdot G_{t_{+}t_{+}}^{r} & 0 & 0\\
\Sigma_{st_{0}}^{r} & 0 & (G_{ss}^{r})^{-1} & 0\\
0 & 0 & 0 & \mathrm{det}\cdot G_{t_{-}t_{-}}^{r}
\end{pmatrix},
\]
where $(G_{\gamma\gamma}^{r})^{-1}=\omega-\omega_{\gamma}-\Sigma_{\gamma
\gamma}^{r}$ and $\mathrm{det}=(G_{t_{0}t_{0}}^{r})^{-1}(G_{ss}^{r}
)^{-1}-(\Sigma_{st_{0}}^{r})^{2}$.

\begin{figure}[tb]
\centering
\psfrag{Ass}{$A_{ss}(\w)$}
\psfrag{Ast0}{$A_{st_0}(\w)$}
\psfrag{At0t0}{$A_{t_0t_0}(\w)$}
\psfrag{w/B}{$\w/B$}
\includegraphics*[width = 0.45 \textwidth]{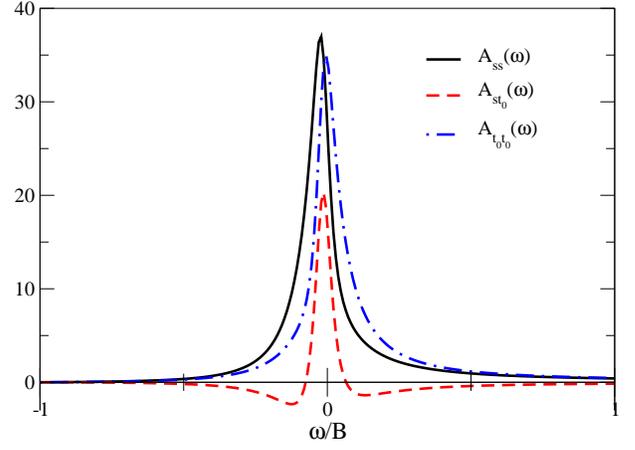}
\caption{(color online) 
%IMPROVE: mention negative values
Spectral functions $A_{ss}, A_{st_0}$ and $A_{t_0t_0}$
versus the frequency $\w/B$
for an exchange spin interaction $K$ of
the order of the level broadening, $K=0.05 \sim \Gamma_{st_0}$.
Further parameters are $B = 1.0$, $g_L = 0.1$, $g_R = 0.2$ and $T = 0.001$.
The off-diagonal spectral function $A_{st_0}$ has total spectral weight zero
and thus contains negative values.
}
\label{fig:Asto_new}
\end{figure}

To lowest order in the coupling to the leads the retarded self energy is of
$\mathcal{O}(g^{2})$,
% CHANGE 
%, and the offdiagonal Green's function is of
%$\mathcal{O}(g^{2}/K)$. 
and the off-diagonal Green's function is given by
$\Sigma_{st_0}^r/\mathrm{det}$. 
Since the total spectral weight of the off-diagonal
spectral function vanishes,  the function changes sign. It can be shown that
it is approximately given by the difference of two Lorentzians, see
illustration in Fig.~\ref{fig:Asto_new}. Since $A_{st_{0}}(\epsilon)$ is not
characterized by a single peak, integration over $\epsilon$ of a product of
$A_{st_{0}}(\epsilon)$ and a more slowly varying function $B(\epsilon)$ may
not be approximated as usual by taking $B(\epsilon)$ at the position of the
peak. 

In order to avoid having to deal with a not positive definite spectral function
we may diagonalize the retarded Green's function matrix. The transformed
matrix is given by
\begin{align*}
(U^{r})_{\gamma_{1}\gamma}^{-1}\mathcal{G}_{\gamma\gamma^{\prime}}^{r}
U_{\gamma^{\prime}\gamma_{2}}^{r}  &  =\tilde{\mathcal{G}}_{\gamma_{1}\gamma_{2}}^{r},\\
\mathrm{and}\qquad\tilde{\mathcal{G}}_{\gamma_{1}\gamma_{2}}^{r}  &  =
\begin{pmatrix}
G_{11}^{r} & 0 & 0 & 0\\
0 & G_{t_{+}t_{+}}^{r} & 0 & 0\\
0 & 0 & G_{22}^{r} & 0\\
0 & 0 & 0 & G_{t_{-}t_{-}}^{r}\\
\end{pmatrix},
\end{align*}
where $G_{11}^{r}=1/(\omega-\omega_{1}^{r})$ and $G_{22}^{r}=1/(\omega
-\omega_{2}^{r})$ with
\begin{align*}
\omega_{1/2}^{r}= & - \frac{1}{2}\left(  
 \w_{t_0} + \Sigma^r_{t_{0}t_{0}}(\omega)
+ \w_s + \Sigma^r_{ss}(\omega)\right)  \\
&  \pm\frac{1}{2}
    \sqrt{\left(  \w_{t_0} + \Sigma^r_{t_0t_0}(\omega) 
                 - \w_s - \Sigma_{ss}^{r}(\omega) \right)^{2}
                 +4\,\left(  \Sigma_{st_{0}}^{r}\right)  ^{2}}.
\end{align*}
This rotation is important only in the case when $\Sigma_{st_{0}}^{r}$ becomes
of the same order of magnitude as the first term in the square root, which is
proportional to the singlet-triplet splitting 
%OLD: $\w_{s} - \w_{t_0} = K$. 
$\w_{t_0} - \w_{s} = K$. 
The transformation matrix is given by
\[
U^{r}=%
\begin{pmatrix}
x_{1}^{r} & 0 & -x_{2}^{r} & 0\\
0 & 1 & 0 & 0\\
x_{2}^{r} & 0 & x_{1}^{r} & 0\\
0 & 0 & 0 & 1
\end{pmatrix}
\]
with the already normalized value of $x_{1}^{r}=(x^{\prime})_{1}^{r}
/\mathcal{N}^{r}$ and $x_{2}^{r}=(x^{\prime})_{2}^{r}/\mathcal{N}^{r}$
where $\mathcal{N}^{r}=\sqrt{\left(  (x^{\prime})_{1}^{r}\right)  ^{2}+\left(
(x^{\prime})_{2}^{r}\right)^{2}}$ and
\begin{align*}
%OLD: (x^{\prime})_{1}^{r}=  &  \frac{1}{2}\left(  (G_{t_{0}t_{0}}^{r}(\omega))^{-1}-(G_{ss}^{r}(\omega))^{-1}\right) \\
%&  +\frac{1}{2}\sqrt{\left(  (G_{t_0t_0}^{r}(\omega))^{-1}-(G_{ss}^{r}(\omega))^{-1}\right)  ^{2}+4\,\left(  \Sigma_{st_{0}}^{r}\right)  ^{2}},\\
(x^{\prime})_{1}^{r}=  &  \frac{1}{2}\left(  (G_{ss}^{r}(\omega
))^{-1}-(G_{t_0t_0}^{r}(\omega))^{-1}\right) \\
&  +\frac{1}{2}\sqrt{\left(  (G_{ss}^{r}(\omega))^{-1}-(G_{t_0t_0}
^{r}(\omega))^{-1}\right)  ^{2}+4\,\left(  \Sigma_{st_{0}}^{r}\right)  ^{2}},\\
(x^{\prime})_{2}^{r}=  &  -\Sigma_{st_{0}}^{r}.
\end{align*}
The transformation matrix $U^{r}$ has complex valued elements and its inverse
is equal to its transpose. 
For $K=0$, we find $x_{1}^r=-x_{2}^r=1/\sqrt{2}$ 
and the eigenstates of the system are given by the product states of the
double quantum dot system, 
$|1\rangle=\frac{1}{\sqrt{2}} (|t_{0}\rangle + |s\rangle)$ and
$|2\rangle=\frac{1}{\sqrt{2}} (|t_{0}\rangle-|s\rangle)$.
%OLD: referee remark: 1/2 instead of 1/\sqrt{2}
%$|1\rangle=\frac{1}{\sqrt{2}}|t_{0}+s\rangle$ and
%$|2\rangle=\frac{1}{\sqrt{2}}|t_{0}-s\rangle$.

The transformation of the advanced Green's function
is different,
\[
\left(  U^{a}\right)  ^{-1}\mathcal{G}^{a}U^{a}=\tilde{\mathcal{G}}^{a},
\]
where the transformation matrix $U^{a}$ is defined in analogy to the retarded
Green's function with $r$ replaced by $a$. 
The transformation matrix $U^a$ is related to $U^r$ by
$\left[ U^a \right]^\dag = \left[ U^r\right]^{-1}$, so that
$[ \tilde{\mathcal{G}}^r ]^\dag = \tilde{\mathcal{G}}^a$.
Since
$\widetilde{G}_{11}^{r},\widetilde{G}_{22}^{r}$ are characterized by a single
pole, the corresponding spectral functions show a single sharp peak and the
commutation relations of the boson operators guarantee the integrated weight
unity. 
\begin{figure}[tb]
\centering
\psfrag{A11}{$\widetilde{A}_{11}(\w)$}
\psfrag{A22}{$\widetilde{A}_{22}(\w)$}
\psfrag{w/B}{$\w/B$}
\includegraphics*[width = 0.45 \textwidth]{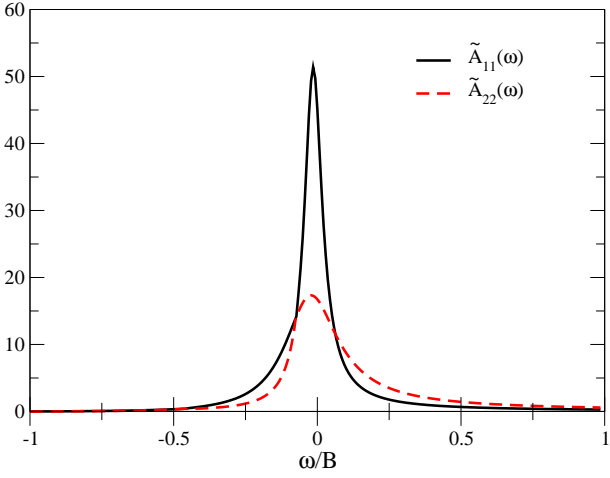}
\caption{(color online) 
Spectral functions $\widetilde{A}_{11}(\omega)$ 
and $\widetilde{A}_{22}(\omega)$ 
versus the frequency $\w/B$ in the rotated space for the same parameter set
as in Fig.~\ref{fig:Asto_new}.}
\label{fig:A11_A22_new}
\end{figure}
In Fig.~\ref{fig:A11_A22_new} 
we show a typical example of the spectral functions
$\widetilde{A}_{ii}(\omega)= - 2 \operatorname{Im}$ $\widetilde
{G}_{ii}^{r}$, $i=1,2$ .

\subsection{Lesser Green's function}
\label{sec:Glesser}

As mentioned above, the lesser Green's function has to be calculated
self-consistently, e.g. in lowest order perturbation theory, thus in this
model to second order.

The lesser components of the Dyson equation are given by one of the equations
\begin{align*}
\left(  \mathcal{G}^{r}\right)  ^{-1}\, \mathcal{G}^{<} 
&= \Sigma^{<}\,\mathcal{G}^{a},\\
\mathcal{G}^{<}\left(  \mathcal{G}^{a}\right)  ^{-1} 
&= \mathcal{G}^{r}\,\Sigma^{<},
\end{align*}
where we neglected the boundary terms. 
%For the following we use the difference of the two latter equations, 
%thus $\mathcal{G}^<$ obeys 
Subtracting these two equations, $G^<$ is found to obey
the quantum Boltzmann equation~\cite{Haug:96},
\begin{align}
\left(  \mathcal{G}^{r}\right)  ^{-1}\,
\mathcal{G}^{<}-\mathcal{G}^{<}\,\left(  \mathcal{G}^{a}\right)  ^{-1}%
=\Sigma^{<}\,\mathcal{G}^{a}-\mathcal{G}^{r}\,\Sigma^{<}.
\label{eq:dyson_lesser_2}
\end{align}
If we neglect the off-diagonal terms in Eq.~\eqref{eq:dyson_lesser_2} we have
to solve the following equations,
%IMPROVE: correct sign
\begin{align*}
\left[  \left(  G^{r}\right)  ^{-1}
       - \left(  G^{a}\right)  ^{-1}\right]_{\gamma\gamma}%(\omega)
\,G_{\gamma\gamma}^{<}%(\omega) 
&=\Sigma_{\gamma\gamma}^{<}%(\omega)
\,\left[  G^{a}-G^{r}\right]_{\gamma\gamma}%(\omega)\\
\\
\Rightarrow\quad G_{\gamma\gamma}^{<}(\omega) &  =
\frac{\Sigma_{\gamma\gamma}^{<}(\omega)}{\Gamma_{\gamma\gamma}(\omega)}
\,A_{\gamma\gamma}(\omega),
\end{align*}
which is a self-consistency equation for the occupation number $n_{\gamma
}=i\,\Sigma_{\gamma\gamma}^{<}(\omega_{\gamma})/\Gamma_{\gamma\gamma}%
(\omega_{\gamma})$. Since the
broadening of the spectral function $A_{\gamma\gamma}$ is the smallest energy
scale in the problem, we neglect the frequency dependence and consider only
the on-shell occupation numbers. 
In this lowest order approximation the quantum
Boltzmann equation~\eqref{eq:dyson_lesser_2} 
is essentially a rate equation (see appendix~\ref{app:n} for
explicit expressions of the occupation numbers), albeit one which now
includes the off-diagonal components of the density matrix.

We will now demonstrate how this approach fails in the example of the
magnetization. For $K = 0$, the left and the
right spins as defined in Eq.~\eqref{eq:def_Sz}
are good quantum numbers and we can define the 
corresponding magnetizations 
%CHANGE: t_+^\dag t_+ + t_-^\dag t_- instead of t_+^\dag t_+^\dag + t_-^'dag t_-^\dag
\[
M_{L/R}=2\langle S_{L/R}^{z}\rangle=\left\langle t_{+}^{\dag}t_{+}
-t_{-}^{\dag}t_{-}\right\rangle
\]
under the assumption that 
%OLD: $\langle st_{0}^{\dag}  +t_{0}^{\dag}s\rangle=0$.
%NEW:
$\langle s^{\dag} t_{0} +t_{0}^{\dag}s\rangle=0$.
This leads to
\[
M_{L/R}=\frac{Y_{L}^{<}(B)+Y_{R}^{<}(B)-Y_{L}^{<}(-B)-Y_{R}^{<}(-B)}{Y_{L}%
^{<}(B)+Y_{R}^{<}(B)+Y_{L}^{<}(-B)+Y_{R}^{<}(-B)},
\]
from which the magnetization on the left quantum dot is found to depend on the
magnetization of the right quantum dot and vice versa, even though the dots are
completely decoupled. 
For a single Kondo impurity, however, 
the correct result is~\cite{Parcollet:02,Paaske:04} 
\begin{align}
M_{L}=\frac{Y_{L}^{<}(B)-Y^<_{L}(-B)}{Y^<_{L}(B)+Y^<_{L}(-B)}%
\label{eq:ML_K0_correct}.
\end{align}
The off-diagonal components are not important in the case of left-right
symmetry and indeed for $Y_{L}=Y_{R}$ we obtain the correct expression for
$M_{L}$. Also, at $B=0$ and consequently 
%OLD: $M=0$ 
$M_L=0$ 
the two different results
coincide. The difference of the two above results for $M_{L}$ can be traced
back to the unjustified neglect of the off-diagonal average $\langle s^{\dag
}t_{0}+t_{0}^{\dag}s\rangle$. We show in the appendix~\ref{app:K0} 
that including the
effect of off-diagonal terms, by employing the transformation defined above, we
recover the correct result for the magnetization $M_{L}$,
Eq.~\eqref{eq:ML_K0_correct}. From this calculation it is obvious that the
off-diagonal terms contribute to the same order as the diagonal terms. 

In the case of finite exchange interaction $K$ the calculation including the
off-diagonal contributions becomes cumbersome, since one has to solve a
self-consistent system of integral equations. The solution can be much
simplified in the transformed basis introduced above.

To rotate the Quantum Boltzmann equation we multiply
Eq.~\eqref{eq:dyson_lesser_2} with $U^{r}$ from the left and 
$(U^{a})^{-1} = (U^r)^\dag$ from the right. 
Thus we find the transformed quantum Boltzmann equation,
\begin{align}
(\tilde{\mathcal{G}}^{r})^{-1}\,\tilde{\mathcal{G}}^{<}-\tilde{\mathcal{G}%
}^{<}\,(\tilde{\mathcal{G}}^{a})^{-1}=\tilde{\Sigma}^{<}\,\tilde
{\mathcal{G}}^{a}-\tilde{\mathcal{G}}^{r}\,\tilde{\Sigma}^{<}%
\label{eq:dyson_lesser_rot}
\end{align}
where 
$\tilde{\Sigma}^{<}=U^{r}\Sigma^{<}(U^{a})^{-1}=U^{r}\Sigma^{<}(U^{r})^\dag$
and 
$\tilde{\mathcal{G}}^{<}=U^{r}\mathcal{G}^{<}(U^{a})^{-1}
                        =U^{r}\mathcal{G}^{<}(U^{r})^\dag$. 
After the
transformation all retarded and advanced Green's functions are diagonal and we
obtain a single equation for every entry of the lesser Green's function. There
is still a finite off-diagonal element $\tilde{G}_{(1,2)}^{<}$ in the
rotated basis, and the big advantage over the initial formulation is 
the fact that the
spectral functions appearing on the r.h.s. of Eq.~\eqref{eq:dyson_lesser_rot}
are now all positive definite
and may be approximated by delta functions of weight unity. As for the
corresponding real parts of $\tilde{\mathcal{G}}^{a,r}$ we adopt the usual
assumption that after frequency integration they may be neglected. 
From Eq.~\eqref{eq:dyson_lesser_rot}
$\tilde{\mathcal{G}}^{<}$ is then found as a sum of delta functions with
weights $\widetilde{\emph{N}}^{\gamma}$ to be determined self-consistently,%
\[
\,\tilde{\mathcal{G}}^{<}=-i%
%TCIMACRO{\tsum \limits_{\gamma=1,2,t_{+},t_{-}}}%
%BeginExpansion
{\textstyle\sum\limits_{\gamma=1,2,t_{+},t_{-}}}
%EndExpansion
\widetilde{\emph{N}}^{\gamma} 2\pi \delta(\omega-\omega_{\gamma}).
\]
This expression, when substituted into the lesser self energy, leads to a
linear combination of weight factors. The quantum Boltzmann equation reduces
to a set of linear homogeneous equations for $\widetilde{\emph{N}}^{\gamma}$, 
which, together with the normalization condition, 
$s^\dag s + t_0^\dag t_0 + t_+^\dag t_+ + t_-^\dag t_- = 1$, may be solved to
give the weight factors. Note that $\widetilde{\emph{N}}^{\gamma}$ are
matrices in the local Hilbert space defined by states $(1,t_{+},2,t_{-})$
with nonzero components $\widetilde{N}_{t_{+}t_{+}}^\gamma$, 
$\widetilde{N}_{t_{-}t_{-}}^\gamma$, $\widetilde{N}_{11}^\gamma$, 
$\widetilde{N}_{22}^\gamma$, $\widetilde{N}_{12}^\gamma$ and 
$\widetilde{N}_{21}^\gamma$.

\begin{figure}[tb]
\centering
\psfrag{<st0+t0s>}{$\langle s^\dag t_0 + t_0^\dag s\rangle$}
\psfrag{V/B}{$eV_L/B$}
\psfrag{K/Gamma=1}{\small $K/\Gamma_{st_0} \simeq 1$}
\psfrag{K/Gamma=1/5}{\small $K/\Gamma_{st_0} \simeq 0.2$}
\psfrag{K/Gamma=0}{\small $K/\Gamma_{st_0} \simeq 0$}
\includegraphics*[width = 0.45 \textwidth]{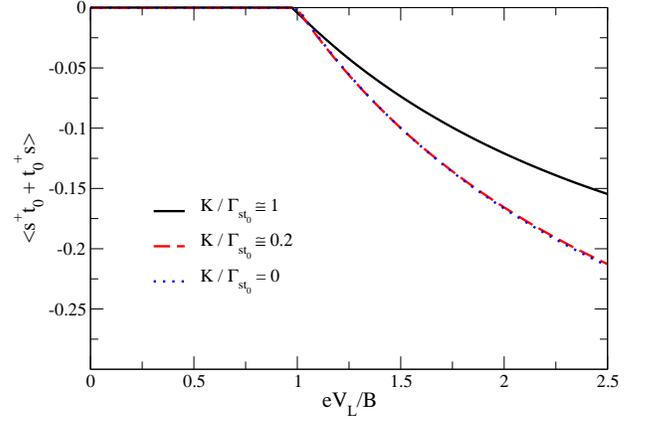}
\caption{(color online) 
The expectation value of $\langle s^\dag t_0 + t_0^\dag s \rangle$, 
i.e.~the contribution of the off-diagonal Green's function 
is of the order of the diagonal contributions
as illustrated here for the cases of $K/\Gamma_{st_0} \simeq 1, 0.2, 0$. 
Further parameters of the plot are 
$B = 1.0$, $g_L = 0.1$, $g_R = 0.2$
and $T = 0.001$.}%
\label{fig:st0t0s}%
\end{figure}

The expectation value $\langle s^{\dag}t_{0}+t_{0}^{\dag}s\rangle$ is
obtained from $G_{st_{0}}^{<}$ as
\begin{align*}
\langle s^{\dag}t_{0}+t_{0}^{\dag}s\rangle & =i\int\frac{d\omega}{2\pi
}[G_{st_{0}}^{<}(\omega)+G_{t_{0}s}^{<}(\omega)]\\
& =i\int\frac{d\omega}{2\pi} 
\left[ \left(x_{1}^{r}x_{2}^{a}+x_2^r x_1^a\right)
       \left(\tilde{G}_{(1,1)}^{<} - \tilde{G}_{(2,2)}^< \right)
\right. \\ &\hspace*{1cm} 
 + \left. 
        \left(x_{1}^{r}x_{1}^{a}-x_2^r x_2^a\right)
        \left(\tilde{G}_{(1,2)}^{<} - \tilde{G}_{(2,1)}^< \right)
\right] \\
& =
       \left(x_{1}^{r}x_{2}^{a}+x_2^r x_1^a\right)
       \left(\widetilde{N}_{11}^{\gamma} - \widetilde{N}_{22}^\gamma \right)
       \\ & \quad
 +     \left(x_{1}^{r}x_{1}^{a}-x_2^r x_2^a\right)
       \left(\widetilde{N}_{12}^{\gamma} - \widetilde{N}_{21}^{\gamma} \right).
\end{align*}
%OLD: and a numerical evaluation of 
A numerical evaluation of 
$\langle s^{\dag}t_{0}+t_{0}^{\dag}s\rangle$ 
is shown in Fig.~\ref{fig:st0t0s}.
For the parameters chosen, it is seen to be 
of the same order of magnitude as for example the non-equilibrium 
magnetization, and is thus comparable to the diagonal
occupation numbers. 
\begin{figure}[tbh]
\centering
\psfrag{ML}{\small $M_L$}
\psfrag{MR}{\small $M_R$}
\psfrag{Mtot}{\small $M_{tot}/2$}
\psfrag{V/B}{$eV_L/B$}
\includegraphics*[width = 0.45 \textwidth]{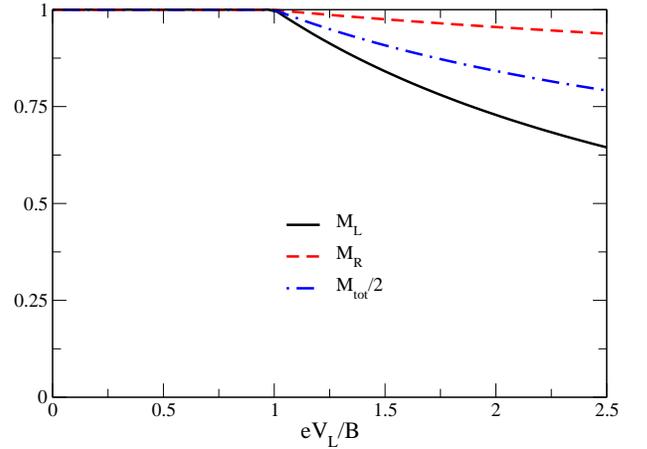}
\caption{(color online) 
The magnetization $M_L$ of the left quantum dot is 
strongly influenced by a voltage applied over the left dot. 
For $K \simeq \Gamma_{st_0}$, the magnetization, $M_R$, 
over the right dot shows only minor deviations from the thermodynamic value.
Neglecting the off-diagonal contributions would give $M_L = M_R = M_{tot}/2$.
Further parameters of the plot are $B = 1.0$, $g_L = 0.1$, $g_R = 0.2$
and $T = 0.001$.}%
\label{fig:M}%
\end{figure}
In Fig.~\ref{fig:M}, we compare the results for the magnetization
with, and without off-diagonal components, in the parameter regime
$K\simeq\Gamma$, where the effect of the off-diagonal components
was argued to become important. The magnetization of the left/right
spin is given by the sum/difference of the total magnetization
and the off-diagonal contributions 
\begin{equation*}
M_{L/R} = 2 \langle \vec{S}_{L/R} \rangle
= \langle (\vec{S}_L + \vec{S}_R) \pm (\vec{S}_L - \vec{S}_R) \rangle,
\end{equation*}
where $\langle \vec{S}_L + \vec{S}_R \rangle
= n_{t_+} - n_{t_-} = M_{tot}/2$
and 
$\langle \vec{S}_L - \vec{S}_R \rangle 
= \langle s^\dag t_0 + t_0^\dag s\rangle$.
For $K = 0$ the 
magnetization of the right quantum dot should not depend
on the voltage applied to the left quantum dot. Therefore
the off-diagonal expectation value, $\langle s^\dag t_0 + t_0^\dag s\rangle$,
has to compensate $100\%$ of the voltage-dependent part of 
$M_{tot}/2$.
For the parameter regime in Fig.~\ref{fig:M}, %and \ref{fig:st0t0s} 
$K \simeq \Gamma_{st_0}$, the compensation from the off-diagonal 
contribution is already of the order of $75\%$.

\section{Calculation of the non-equilibrium current}

The non-linear conductance $dI/dV$ is governed by the voltage-dependence of
the occupation of states. For increasing voltage $V$, 
the conductance will have a
step when the energy supplied by $V$ allows the occupation of an excited
state. The voltage-dependence of the level occupations will change the step to
a cusp, and higher order in the perturbation series will change this to a
logarithmic nonequilibrium Kondo peak, cut-off by
the spin-dependent relaxation rate. 

To second order in the exchange-tunnel coupling 
the current 
%ADD text
through the left quantum dot 
%ADD reference
is given by~\cite{Paaske:04, Meir_Wingreen, Rosch, Thesis}
\begin{align*}
I_L=&-\frac{\pi}{8}\frac{e}{h}g_{12}g_{21} \\
&\int d\omega X_{DQD}^{>}(\omega)\left[
B(\omega-eV_L)-B(\omega+eV_L)\right],
\end{align*}
%OLD: where $X_{DQD}(\omega)$ is the susceptibility of the Double Quantum Dot
%System
where $X^>_{DQD}(\omega)$ is the susceptibility of the double quantum dot
system
\[
X_{DQD}^{>}(\omega)=\sum_{all}\tau_{\sigma^{\prime}\sigma}^{i}%
\tau_{\sigma\sigma^{\prime}}^{j}\int\frac{d\epsilon}{2\pi}\mathrm{Tr}\left[
\mathcal{G}^{<}(\epsilon)T_{L}^{i}\mathcal{G}^{>}(\epsilon+\omega)T_{L}^{j}\right].
\]
This expression has to be modified according to 
the rotation in the basis states.
Therefore we use the representation 
$\mathcal{G}^{>}=\mathcal{G}^{r}-\mathcal{G}^{a}+\mathcal{G}^{<}$, 
together with the fact that terms containing more than one factor of 
$\mathcal{G}^{<}$ are projected
out. The lesser Green's function is given by 
$\mathcal{G}^< = U^a \mathcal{\tilde G}^< (U^r)^{-1}$
and the retarded and advanced Green's functions, 
$\mathcal{G}^{r}$ and $\mathcal{G}^{a}$, 
have to be transformed according to 
$\mathcal{G}^{r/a} = (U^{r/a})^{-1} \mathcal{\tilde G}^{r/a} U^{r/a}$.
The two limiting cases $\Sigma_{st_0} = 0$ and $K = 0$
are discussed in appendix~\ref{app:current}. The
off-diagonal contributions show a significant effect also in
an intermediate regime.

\begin{figure}[t!b]
\centering
\psfrag{V/B}{$eV_L/B$}
\psfrag{dI/dV}{$dI_L/dV_L$}
%\psfrag{with}{\small $\langle s^\dag t_0 + t_0^\dag s\rangle \not= 0$}
\psfrag{K/Gamma=1 without}{%\small $K/\Gamma \approx 1$, 
                           $\langle s^\dag t_0 + t_0^\dag s\rangle = 0$}
\psfrag{K/Gamma=1}{\small $K/\Gamma_{st_0} \simeq 1$}
\psfrag{K/Gamma=1/5}{\small $K/\Gamma_{st_0} \simeq 0.2$}
\psfrag{K/Gamma=0}{\small $K/\Gamma_{st_0} \simeq 0$}
\includegraphics*[width = 0.45 \textwidth]{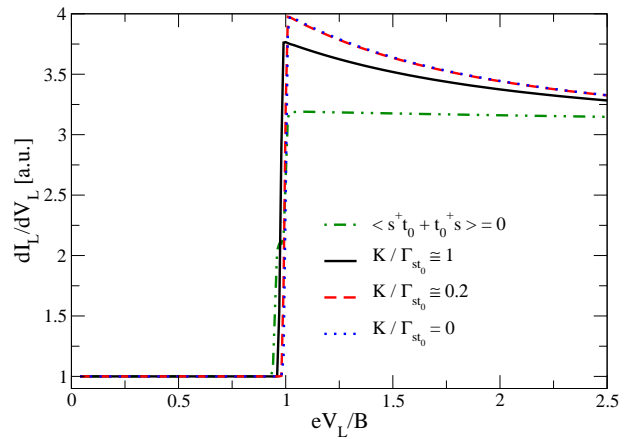}
\caption{(color online) 
%OLD: Current 
Differential conductance (in units of 
$\left(\pi/8\right) \left(e^2/h\right) g_{12} g_{21}$)
in the case of small exchange interaction 
$K/\Gamma_{st_0} \simeq 1, 0.2, 0$. For $K \simeq \Gamma_{st_0}$
the results including or neglecting the off-diagonal contributions
are shown. The further parameters
are chosen identically to the previous figures.}
\label{fig:current}%
\end{figure}

In Fig.~\ref{fig:current} the 
%OLD: current
differential conductance 
is plotted for the parameters
$K/\Gamma_{st_0} \simeq 1, 0.2, 0$, and for $K \simeq \Gamma_{st_0}$
both with and without the correction caused by a
finite expectation value $\langle s^{\dag}t_{0}+t_{0}^{\dag}s\rangle$. 
One observes a significant difference, 
in particular near threshold. Since the current expression
depends sensitively on the non-equilibrium occupation numbers,
the physically relevant results are given only for
the correct occupation numbers. If the finite contribution
of $\langle s^\dag t_0 + t_0^\dag s \rangle$ is neglected,
the current through the left quantum dot
depends on the total magnetization of the double quantum dot system,
such that e.g. a finite voltage on the right quantum dot would
affect the current through the left, although the two quantum dots
are decoupled when $K = 0$, see also discussion 
in the appendix~\ref{app:current}.

\section{Conclusion}

The state of an isolated nanostructure is determined by the specification of
the occupation of the eigenstates of the system. In other words, the density
matrix (the lesser Green's function integrated over frequency) 
of the isolated system is diagonal in the
basis of eigenstates. Coupling of the nanostructure to reservoirs will in
general lead to a change of the density matrix. This change may involve the
appearance of off-diagonal elements in the density matrix. These off-diagonal
elements generically have a more complex frequency dependence than the
diagonal terms. 
While the diagonal terms of the density matrix have a spectral
function characterized by a single narrow peak, and a spectral weight to be
interpreted as the occupation number of the state in question, the spectral
functions of the off-diagonal elements of the density matrix have positive and
negative parts and total spectral weight zero. 
Nonetheless these off-diagonal
terms may be as important as the diagonal ones, as we demonstrate in the
example of a double quantum dot system in a magnetic field. 

%NEW: new paragraph
We have presented
a systematic method of how to deal with this problem, by introducing the two
sets of eigenstates of the retarded and advanced Green's functions. In terms
of these eigenstates the quantum Boltzmann equation may be solved in the usual
approximation of assuming the spectral functions to be delta functions. The
method is generally applicable, but it is demonstrated here 
%NEW:
in the example 
%OLD: for the example 
of a minimal model, where the problem of off-diagonal elements of the density
matrix arises: a double quantum dot system coupled by spin exchange
interaction in a magnetic field.

%NEW:
In this case, the eigenstates of the isolated double-dot system 
are the singlet and triplet states, even for arbitrarily small exchange 
interaction $K$. However, when the leads are coupled to the dot system, 
and K is of order, or less than the level broadening on the dot system 
induced by the leads, the eigenstates of the coupled system - quantum dots 
plus leads - approach the product states of the spins 1/2 of 
the individual dots. The transition in the character of states as 
the exchange coupling $K$ is varied is captured perfectly by the 
representation in the rotated basis proposed here. Our method thus allows 
to avoid the time-consuming numerical solution of the full 
frequency dependent quantum Boltzmann equation. 

The method is quite general and can be applied to a wide range of 
quantum-impurity problems. This is particularly relevant in 
multi-orbital problems such as carbon nanotube quantum dots and 
single-molecule transistors involving smaller conjugated molecules, 
possibly acting as high-spin impurities.

%\bigskip

\section*{Acknowledgments}

We acknowledge discussions with J.~Lehmann.
This work has been supported
by the DFG-Center for Functional Nanostructures
(CFN) at the University of Karlsruhe under project B2.9
(V.K. and P.W.), the Institute for Nanotechnology,
Research Center Karlsruhe (P.W.), and the Danish
Agency for Science, Technology and Innovation (J.~P.).

\appendix

\section{Derivation of the self energy}

\label{sec:app_selfenergy}
%{\it Appendix A}

From Eq.~\eqref{eq:general_selfenergy_1} we find the following self energies
(a common prefactor of -1/16 is implied),
\begin{widetext}
\begin{align}
\Sigma_{ss} (\tau_1, \tau_2) =&
\left( G_{t_0t_0}(\tau_1, \tau_2) + G_{t_+t_+}(\tau_1, \tau_2)
+ G_{t_-t_-}(\tau_1, \tau_2) \right)
%\left( Y_L(\tau_1, \tau_2) + Y_R(\tau_1, \tau_2) \right)
Y_{L+R}(\tau_1, \tau_2) 
\label{eq:Sigma_ss}, \\
\Sigma_{t_0t_0} (\tau_1, \tau_2) =&
\left( G_{ss}(\tau_1, \tau_2) + G_{t_+t_+}(\tau_1, \tau_2)
+ G_{t_-t_-}(\tau_1, \tau_2) \right)
%\left( Y_L(\tau_1, \tau_2) + Y_R(\tau_1, \tau_2) \right)
Y_{L+R}(\tau_1, \tau_2) 
\label{eq:Sigma_t0t0},\\
\Sigma_{t_\pm t_\pm} (\tau_1, \tau_2) =&
\left( G_{ss}(\tau_1, \tau_2) + G_{t_0t_0}(\tau_1, \tau_2)
+ G_{t_\pm t_\pm}(\tau_1, \tau_2) \right)
%\left( Y_L(\tau_1, \tau_2) + Y_R(\tau_1, \tau_2) \right)  \nonumber \\
Y_{L+R}(\tau_1, \tau_2)  \nonumber \\
&\mp \left( G_{st_0}(\tau_1, \tau_2) + G_{t_0s}(\tau_1, \tau_2) \right)
Y_{L-R}(\tau_1, \tau_2) 
\label{eq:Sigma_tptp}, 
%\Sigma_{t_-t_-} (\tau_1, \tau_2) =&
%\left( G_{ss}(\tau_1, \tau_2) + G_{t_0t_0}(\tau_1, \tau_2)
%+ G_{t_-t_-}(\tau_1, \tau_2) \right)
%\left( Y_L(\tau_1, \tau_2) + Y_R(\tau_1, \tau_2) \right) \nonumber \\
%&+ \left( G_{st_0}(\tau_1, \tau_2) + G_{t_0s}(\tau_1, \tau_2) \right)
%\left( Y_L(\tau_1, \tau_2) - Y_R(\tau_1, \tau_2) \right)
%\label{eq:Sigma_tmtm}
\intertext{and}
\Sigma_{st_0} (\tau_1, \tau_2) =&
\left( G_{t_-t_-}(\tau_1, \tau_2) - G_{t_+t_+}(\tau_1, \tau_2) \right)
%\left( Y_L(\tau_1, \tau_2) - Y_R(\tau_1, \tau_2) \right) \nonumber \\
Y_{L-R}(\tau_1, \tau_2)  %\nonumber \\
+ G_{t_0s}(\tau_1, \tau_2)
Y_{L+R}(\tau_1, \tau_2),
%    \label{eq:Sigma_st0}\\
\\
\Sigma_{t_0s} (\tau_1, \tau_2) =&
\left( G_{t_-t_-}(\tau_1, \tau_2) - G_{t_+t_+}(\tau_1, \tau_2) \right)
%\left( Y_L(\tau_1, \tau_2) - Y_R(\tau_1, \tau_2) \right) \nonumber \\
Y_{L-R}(\tau_1, \tau_2)  %\nonumber \\
+ G_{st_0}(\tau_1, \tau_2)
%\left( Y_L(\tau_1, \tau_2) + Y_R(\tau_1, \tau_2) \right)
Y_{L+R}(\tau_1, \tau_2) 
\label{eq:Sigma_t0s}.
\end{align}
\end{widetext}

\section{Solution of the quantum Boltzmann equation
without off-diagonals}
\label{app:n}

%{\it Appendix B}

If we neglect the off-diagonal elements in Eq.~\eqref{eq:dyson_lesser_2} we
have to solve the following equations,
\begin{align*}
%\left[  \left(  G^{r}\right)  ^{-1}-\left(  G^{a}\right)  ^{-1}\right]
%_{\gamma\gamma}(\omega)\,G_{\gamma\gamma}^{<}(\omega) &  =\Sigma_{\gamma
%\gamma}^{<}(\omega)\,\left[  G^{a}-G^{r}\right]  _{\gamma\gamma}(\omega)\\
%\Rightarrow\quad 
G_{\gamma\gamma}^{<}(\omega) &  = \frac{\Sigma_{\gamma\gamma}^{<}%
(\omega)}{\Gamma_{\gamma\gamma}(\omega)}\,A_{\gamma\gamma}(\omega).
\end{align*}
which corresponds to a self-consistency equation for the occupation number
$n_{\gamma}=i\,\Sigma_{\gamma\gamma}^{<}(\omega_{\gamma})/\Gamma_{\gamma\gamma}%
(\omega_{\gamma})$, when neglecting the frequency dependence 
and considering only the on-shell occupation numbers. 
All other quantities will change on a
larger energy scale than the broadening of the spectral function $A_{\gamma}$,
therefore we are allowed to approximate it by a $\delta$-function. 
This simplifies the frequency integration in the self energies, 
leading to a set of
homogeneous %OLD: linear equations for the $n_{\gamma}$.
linear equations for $n_{\gamma}$.
These equations are closed
by imposing the normalization condition 
${\textstyle\sum\limits_{\gamma}}
n_{\gamma}=1$. The solution is
\[
n_{\gamma}=\frac{\mathcal{N}_{\gamma}}{\mathcal{N}_{s}+\mathcal{N}_{t_{0}%
}+\mathcal{N}_{t_{+}}+\mathcal{N}_{t_{-}}},
\]
where
\begin{align*}
Z_{t_{0}} &  =Y_{L+R}^{<}(-K)+Y_{L+R}^{<}(B)+Y_{L+R}^{<}(-B),\\
Z_{t_{\pm}} &  =Y_{L+R}^{<}(-K\pm B)+Y_{L+R}^{<}(\pm B),
\end{align*}
and \begin{widetext}
\begin{align*}
\mathcal{N}_s =& Z_{t_0} Z_{t_+} Z_{t_-}
- \left( Z_{t_-} + Z_{t_+} \right) Y^<_{L+R}(B) Y^<_{L+R}(- B), \\
\mathcal{N}_{t_0} =& Z_{t_+} Z_{t_-} Y_{L+R}^<(K)
+ Z_{t_+} Y_{L+R}^<(- B) Y_{L+R}^<(K + B)
+ Z_{t_-} Y_{L+R}^<(B) Y_{L+R}^<(K - B), \\
\mathcal{N}_{t_\pm} =& Z_{t_\mp} Z_{t_0} Y_{L+R}^<(K \mp B)
+ Z_{t_\mp} Y_{L+R}^<(\mp B) Y_{L+R}^<(K) \\
&+ Y_{L+R}^<(\mp B) Y_{L+R}^<(\mp B)  Y_{L+R}^<(K \pm B)
- Y_{L+R}^<(\mp B) Y_{L+R}^<(\pm B)  Y_{L+R}^<(K \mp B).
\end{align*}
\end{widetext}

\section{Quantum Boltzmann equation in the limiting case $K = 0$}
\label{app:K0}
%{\it Appendix C}

In the special case of $K=0$ the $ss$ and $t_{0}t_{0}$ components 
have the same energy, $\omega_{s}=\omega_{t_{0}}=0$. 
Their zeroth order contributions are identical and they will be the same in
every following order of the calculation. 
The same argument holds for $st_{0}$ and $t_{0}s$.

Thus we have to solve only the two equations following from
Eq.~\eqref{eq:dyson_lesser_2}
\begin{align*}
&  \left(  G_{ss}^{r}\right)  ^{-1}\,G_{ss}^{<}-G_{ss}^{<}\,\left(  G_{ss}%
^{a}\right)  ^{-1}+\left(  G_{st_0}^{r}\right)  ^{-1}\,G_{t_0s}^{<}-G_{st_0}%
^{<}\,\left(  G_{t_0s}^{a}\right)  ^{-1}\\
&  =\Sigma_{ss}^{<}\,G_{ss}^{a}-G_{ss}^{r}\,\Sigma_{ss}^{<}+\Sigma_{st_0}%
^{<}\,G_{t_0s}^{a}-G_{st_0}^{r}\,\Sigma_{t_0s}^{<},\\[0.5cm]
&  \left(  G_{ss}^{r}\right)  ^{-1}\,G_{st_0}^{<}-G_{ss}^{<}\,\left(  G_{st_0}%
^{a}\right)  ^{-1}+\left(  G_{st_0}^{r}\right)  ^{-1}\,G_{t_0t_0}^{<}-G_{st_0}%
^{<}\,\left(  G_{t_0t_0}^{a}\right)  ^{-1}\\
&  =\Sigma_{ss}^{<}\,G_{st_0}^{a}-G_{ss}^{r}\,\Sigma_{st_0}^{<}+\Sigma_{st_0}%
^{<}\,G_{t_0t_0}^{a}-G_{st_0}^{r}\,\Sigma_{t_0t_0}^{<}.
\end{align*}
The sum and difference of these two equations, using $G_{st_{0}}=G_{t_{0}s}$
and $G_{ss}=G_{t_{0}t_{0}}$ in the special case of $K=0$, are
\begin{align*}
&  \left[  \left(  G_{ss}^{r}\right)  ^{-1}\pm\left(  G_{st_0}^{r}\right)
^{-1}-\left(  G_{ss}^{a}\right)  ^{-1}\mp\left(  G_{st_0}^{a}\right)
^{-1}\right]  \,\left[  G_{ss}^{<}\pm G_{st_0}^{<}\right]  \\
&  =\left[  \Sigma_{ss}^{<}\pm\Sigma_{st_0}^{<}\right]  \,\left[  G_{ss}^{a}\pm
G_{st_0}^{a}-G_{ss}^{r}\mp G_{st_0}^{r}\right].
\end{align*}
%OLD: For zero spin interaction we have to do perturbation theory in the rotated
For zero exchange interaction $K = 0$
we have to do perturbation theory in the rotated
eigenspace
\[%
\begin{pmatrix}
|1\rangle\\
|2\rangle
\end{pmatrix}
=\frac{1}{\sqrt{2}}%
\begin{pmatrix}
1 & 1\\
-1 & 1
\end{pmatrix}%
\begin{pmatrix}
|s\rangle\\
|t_{0}\rangle
\end{pmatrix},
\]
where the new states $|1\rangle$ and $|2\rangle$ 
correspond to product states and
are orthonormal. Again using the strongly peaked nature of the spectral
functions we find a set of linear equations for the occupation numbers, which,
amended with the normalization condition, give the solution
\begin{align*}
n_{1} &  =\frac{Y^<_{L}(-B)Y^<_{R}(B)}{\left(  Y_{L}^{<}(B)+Y_{L}^{<}(-B)\right)
\left(  Y_{R}^{<}(B)+Y_{R}^{<}(-B)\right)  },\\
n_{2} &  =\frac{Y^<_{L}(B)Y^<_{R}(-B)}{\left(  Y_{L}^{<}(B)+Y_{L}^{<}(-B)\right)
\left(  Y_{R}^{<}(B)+Y_{R}^{<}(-B)\right)  },\\
n_{t_{\pm}} &  =\frac{Y^<_{L}(\mp B)Y^<_{R}(\mp B)}{\left(  Y_{L}^{<}(B)+Y_{L}%
^{<}(-B)\right)  \left(  Y_{R}^{<}(B)+Y_{R}^{<}(-B)\right)  }.
\end{align*}
The discussion in the main text shows, that the contribution of the
off-diagonal elements
\[
\langle s^{\dag}t_{0}+t_{0}^{\dag}s\rangle=n_{1}-n_{2}%
\]
is of the same order as the diagonal contributions like $\langle t_{-}^{\dag
}t_{-}-t_{+}^{\dag}t_{+}\rangle$. 
Please note, that within this calculation the occupation numbers,
$n_{1}  = n_{L\uparrow} n_{R\downarrow}$,
are given as product states. Although the two quantum dots are decoupled in
the case of $K = 0$, the solution for the occupation numbers of the product
states contains information about the left and right quantum dot simultaneously.

Finally, we find for the magnetization
\begin{align*}
M_{L} =  &  \left\langle t_{+}^{\dag}t_{+} - t_{-}^{\dag}t_{-} \right\rangle
\\
&  + \left\langle (s + t_{0})^{\dag}(s + t_{0}) - (t_{0} - s)^{\dag}(t_{0} - s) \right\rangle \\
=  &  n_{L\uparrow} n_{R\uparrow} - n_{L\downarrow} n_{R\downarrow} +
n_{L\uparrow} n_{R\downarrow} - n_{L\downarrow} n_{R\uparrow}\\
=  &  n_{L\uparrow} - n_{L\downarrow}%
= \frac{ Y^<_{L}(B) - Y^<_{L}(- B)}{ Y^<_{L}(B) + Y^<_{L}(- B) } .
\end{align*}

\section{Explicit current expressions}
\label{app:current}

The general expression for the current will not be given here,
but we like to discuss briefly two limiting cases.
If the off-diagonal contributions are zero, for example
for $B = 0$ or left-right symmetry, the current is
given by
\begin{align}
I_L=(2\pi)^{2}\frac{e}{h}\frac{1}{8}g_{12}^{2} &  \left[  3eV_L+\left(
n_{s}-n_{t_{0}}\right)  F_{3}(K,V_{L})\right.  \nonumber \\
&  \left.  +\left(  n_{s}-n_{t_{-}}\right)  F_{3}(K+B,V_{L})\right.  \nonumber \\
&  \left.  +\left(  n_{s}-n_{t_{+}}\right)  F_{3}(K-B,V_{L})\right.  \nonumber \\
&  \left.  +\left(  n_{t_{0}}-n_{t_{-}}-n_{t_{0}}+n_{t_{+}}\right)
F_{3}(B,V_{L})\right]
\label{eq:current_K0},
\end{align}
where the function
\begin{align*}
F_{3}(x, V) =  &  \frac{1}{2} (x - eV) \coth\left(  \frac{1}{2} \beta(x - eV)
\right) \\
&  - \frac{1}{2} (x + eV) \coth\left(  \frac{1}{2} \beta(x + eV) \right)
\end{align*}
is asymmetric in the voltage $V$ and also in the energy $x$.
The function $F_3(x, V)$ describes the non-linear behavior
of the differential conductance.

In contrast, for the case of $K=0$, the current is given by
\begin{align*}
I_L &  =(2\pi)^{2}\frac{e}{h}\frac{1}{8}g_{12}^{2} \\
&  \left[  3eV_L+2\left( n_{t_{+}}-n_{t_{-}}+n_{1}-n_{2}\right)  
    F_{3}(B,V_{L})\right].
\end{align*}
This result is identical to the current
for a single quantum dot in non-equilibrium~\cite{Paaske:04},
since $n_{t_+} - n_{t_-} + n_1 - n_2 = M_L$. The
current depends only on the magnetization of the left
quantum dot. Taking the limit of $K \to 0$ 
in the expression~\eqref{eq:current_K0},
the current is proportional to the total magnetization
$(n_{t_+} - n_{t_-})$ and thus dependent on properties
of the right quantum dot, which is obviously wrong.
Please compare with the discussion of the magnetization 
in section~\ref{sec:Glesser}.


\begin{thebibliography}{99}

\bibitem {Rammer:86}J.~Rammer and H.~Smith, Rev.~Mod.~Phys. \textbf{58}, 323
(1986).

\bibitem {Haug:96}H.~Haug and A.-P.~Jauho, 
\textit{Quantum Kinetics in Transport and Optics of Semiconductors}, 
Springer Series in Solid-State Science, Springer-Verlag, (1996).

\bibitem {Parcollet:02}O.~Parcollet, and C.~Hooley, Phys. Rev. B \textbf{66},
085315 (2002).

\bibitem {Paaske:04}J.~Paaske, A.~Rosch, and P.~W\"olfle, Phys. Rev. B
\textbf{69}, 155330 (2004).

\bibitem{Bloch:53} R.~K.~Wangsness and F.~Bloch, 
                Phys. Rev. \textbf{89}, 728 (1953).

\bibitem{Redfield:57} A.~G.~Redfield, IBM J. Res. Dev. \textbf{1}, 19 (1957).

\bibitem{Blum} K. Blum, \textit{Density Matrix Theory and Applications} 
Plenum, New York, (1996). 

\bibitem{Slichter} C.~P.~Slichter \textit{Principles of Magnetic Resonance},
                   Springer-Verlag, (1990).

\bibitem{Engel} H.-A.~Engel and D.~Loss, 
                Phys. Rev. Lett. \textbf{86}, 4648 (2001).

\bibitem{Lehmann} J.~Lehmann, A.~Gaito-Ari{\~n}o, E.~Coronado and D.~Loss,
                  Nature Nanotech. \textbf{2}, 312 (2007).

\bibitem {Sachdev:90}S.~Sachdev and R.~N.~Bhatt, Phys. Rev. B \textbf{41} 9323 (1990).

\bibitem {Abrikosov:65}A.~A.~Abrikosov, Physics \textbf{2}, 5 (1965).

\bibitem {Koerting:07}V.~Koerting, W.~W{\"o}lfle and J.~Paaske,
Phys.~Rev.~Lett. 99, 036807 (2007).

\bibitem {Meir_Wingreen}Y.~Meir and N.~S.~Wingreen, Phys.~Rev.~Lett.
\textbf{68}, 2512 (1992).

%\bibitem {master_frederikson}find reference

\bibitem {Rosch}A. Rosch, J. Paaske, J. Kroha, and P. W\"{o}lfle, Phys.Rev.
Lett. 90, 076804 (2003); J. Phys. Soc. Jpn. 74, 118 (2005).

\bibitem{Thesis} V.~Koerting, \textit{Non-equilibrium Electron Transport through a Double Quantum Dot System} PhD thesis, University of Karlsruhe, December 2007.
\end{thebibliography}
\end{document}